\documentclass[]{emulateapj}
\usepackage{float}
\bibpunct{(}{)}{;}{a}{}{,}

\slugcomment{accepted by ApJL, August 18, 2015}
\shorttitle{Another star with $\mbox{[Fe/H]} = -5.0$}
\shortauthors{Frebel et al.}

\begin{document}

\title{SD~1313$-$0019 -- Another second generation star with
  $\mbox{[Fe/H]} = -5.0$, \\observed 
  with the Magellan Telescope\altaffilmark{*}}

\author{
Anna Frebel\altaffilmark{1},
Anirudh Chiti\altaffilmark{1},
Alexander P. Ji\altaffilmark{1},
Heather R. Jacobson\altaffilmark{1},
and Vinicius M. Placco\altaffilmark{2}}

\altaffiltext{*}{This paper includes data gathered with the 6.5 meter
  Magellan Telescopes located at Las Campanas Observatory, Chile.}

\altaffiltext{1}{Department of Physics and Kavli Institute for
  Astrophysics and Space Research, Massachusetts Institute of
  Technology, Cambridge, MA 02139, USA}

\altaffiltext{2}{Department of Physics and JINA Center for the
  Evolution of the Elements, University of Notre Dame, Notre Dame, IN
  46556, USA}

\begin{abstract}
We present a Magellan/MIKE high-resolution ($R\sim35,000$) spectrum of
the ancient star SD~1313$-$0019 which has an iron abundance of
$\mbox{[Fe/H]} = -5.0$, paired with a carbon enhancement of
$\mbox{[C/Fe]} \sim 3.0$.  The star was initially identified by
Allende Prieto et al. in the BOSS survey. Its medium-resolution
spectrum suggested a higher metallicity of $\mbox{[Fe/H]}= -4.3$ due
to the Ca\,II\,K line blending with a CH feature which is a common
issue related to the search for the most iron-poor stars. This star
joins several other, similar stars with $\mbox{[Fe/H]} \lesssim -5.0$
that all display a combination of low iron and high carbon
abundances. Other elemental abundances of SD~1313$-$0019 follow that
of more metal-rich halo stars. Fitting the abundance pattern with
yields of Population\,III supernova suggests that SD~1313$-$0019
had only one massive progenitor star with 20 - 30M$_{\odot}$ that must
have undergone a mixing and fallback episode. Overall, there are now
five stars known with $\mbox{[Fe/H]} \lesssim -5.0$ (1D LTE
abundances). This ever-increasing population of carbon-rich,
iron-deficient stars can potentially constrain nucleosynthesis in
Population\,III stars and their supernova explosions, the formation
mechanisms of the first low mass stars, and the nature of the first
galaxies.
\end{abstract}


\section{Introduction}

The chemical abundances of the most iron-poor stars provide a
record of the nucleosynthesis yields of the first stars that formed
in the Universe. By now, $\sim20$ Milky Way halo stars are known with
$\mbox{[Fe/H]} \lesssim -4.0$ (see e.g., \citealt{fn15,placco15}) of
which five have $\mbox{[Fe/H]} \lesssim -5.0$
\citep{HE0107_Nature,HE1327_Nature,caffau11, keller14,bonifacio15}.
They are thought to be second-generation stars from the early Universe
with just one massive Population\,III (hereafter Pop\,III) first
star progenitor that produced the chemical elements that we can still
observe in their stellar atmospheres today.

These stars have been used to constrain the properties of the first
stars (e.g., \citealt{limongi_he0107,meynet06, heger10, tominaga14}),
the formation sites of early low-mass stars (e.g.,
\citealt{karlsson13,cooke14, smith15}) and their formation mechanisms
\citep{dtrans,schneider12b, ji14}. Moreover, they provide key
information on the production of the first elements that ushered in
the chemical evolution of the universe \citep{fn15}.  Clearly, these
stars are at the heart of stellar archaeology and of vital importance
to near-field cosmology. Accordingly, several efforts are underway to
uncover additional most iron-poor stars \citep{keller07, caffau13,
  schlaufman14, li15a}, reflecting the wide ranging interest in these
ancient stars. This follows the HK survey \citep{BPSII}, the
Hamburg/ESO survey \citep{hes4, frebel_bmps}, and the Sloan Digital
Sky Survey (SDSS) and Sloan Extension for Galactic Understanding and
Exploration (SEGUE) survey \citep{aoki13, caffau13, bonifacio15}.

We report the detailed chemical abundances of yet another hyper
iron-poor star with $\mbox{[Fe/H]} = -5.0$. The low-metallicity nature
of SDSS~J131326.89$-$001941.4 (hereafter SD~1313$-$0019) was initially
recognized by \citet{allendeprieto15} based on an medium-resolution
SDSS-BOSS (Baryonic Oscillation Spectroscopic Survey) spectrum with
$R\sim2,000$. The analysis of this spectrum showed the star to have
$\mbox{[Fe/H]} \lesssim -4.3$ and $\mbox{[C/Fe]} = 2.5$, making
SD~1313$-$0019 an ideal target for high-resolutions spectroscopic
followup observations to confirm medium-resolution results and
determine additional chemical abundances.

\section{Observations}\label{sec:obs}

We observed SD~1313$-$0019 (R.A. = 13:13:26.89, Dec. = $-$00:19:41.4,
$V=16.9$) with the MIKE spectrograph \citep{mike} on the Magellan-Clay
telescope at Las Campanas Observatory on June 16 and 17, 2015. A
$0\farcs7$ slit yields a high spectral resolution of $\sim28,000$ in
the red and $\sim35,000$ in the blue wavelength regime. The total
exposure time was 3\,h, although the seeing degraded during the second
half of the observations.

\begin{figure}[!ht]
 \begin{center}
  \includegraphics[clip=true,width=7cm, bbllx=44, bblly=123,
    bburx=545,
    bbury=458]{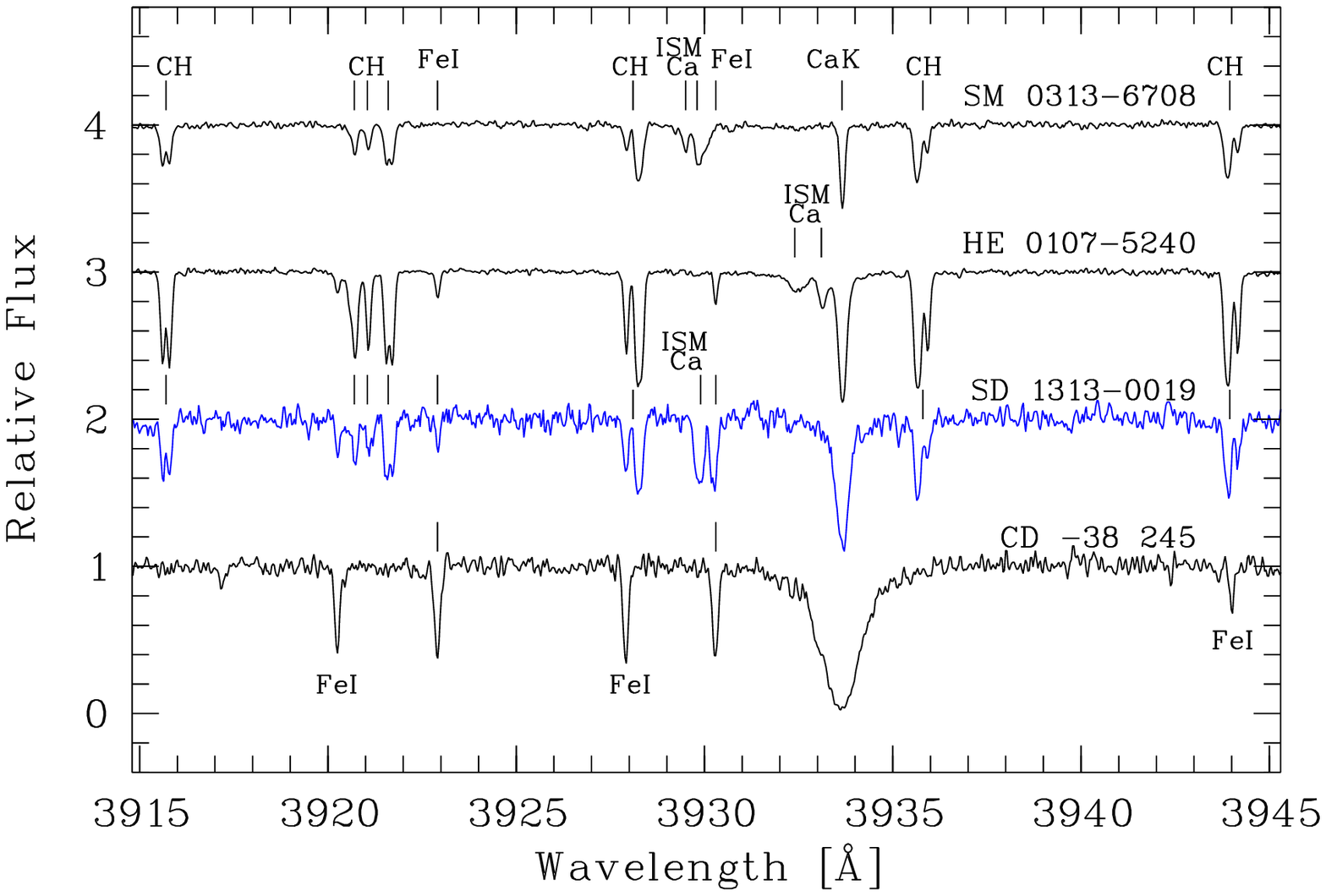}\\ \includegraphics[clip=true,width=7cm,
    bbllx=44, bblly=124, bburx=545,
    bbury=458]{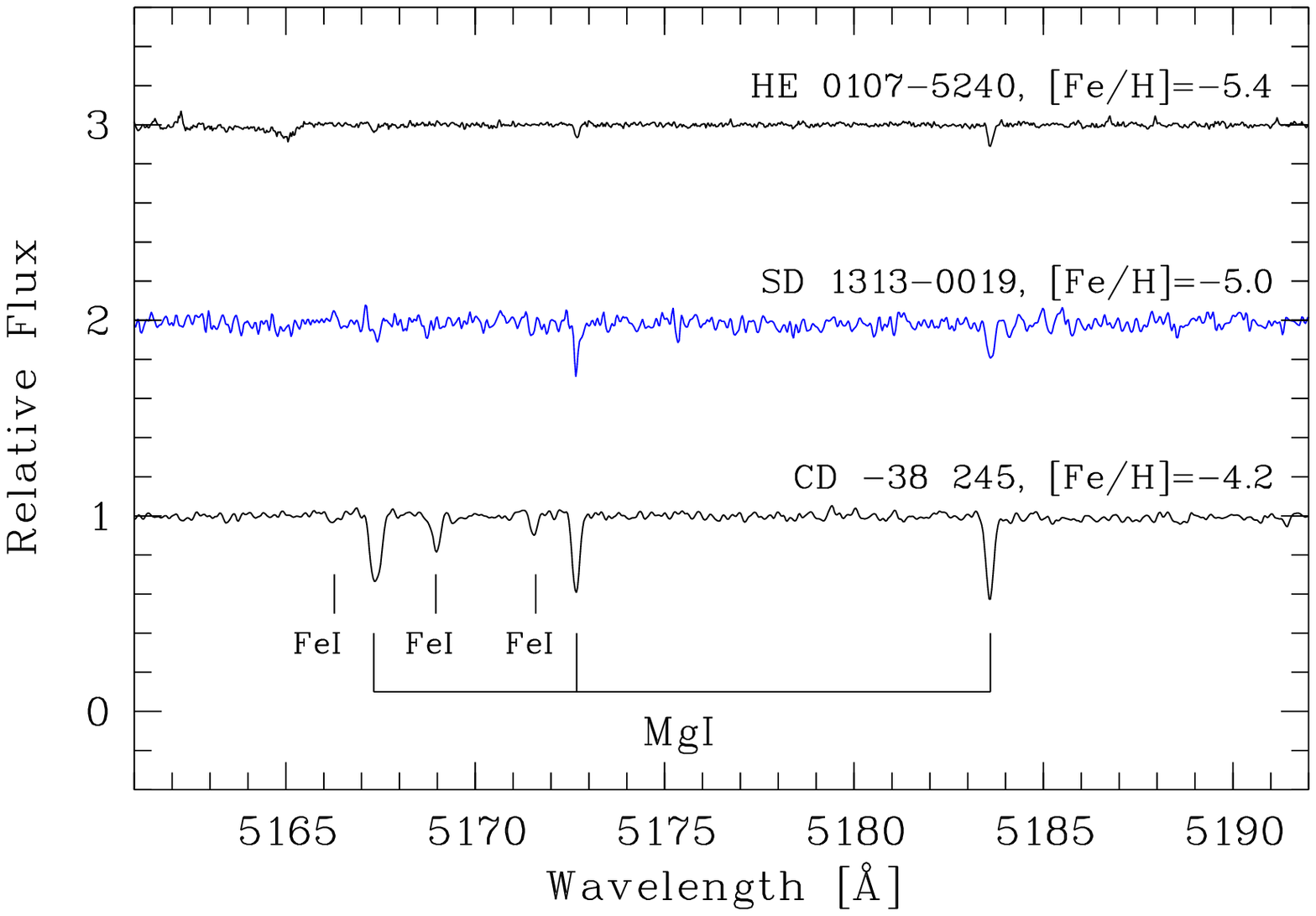}\\ \includegraphics[clip=true,width=7cm,
    bbllx=40, bblly=124, bburx=558,
    bbury=458]{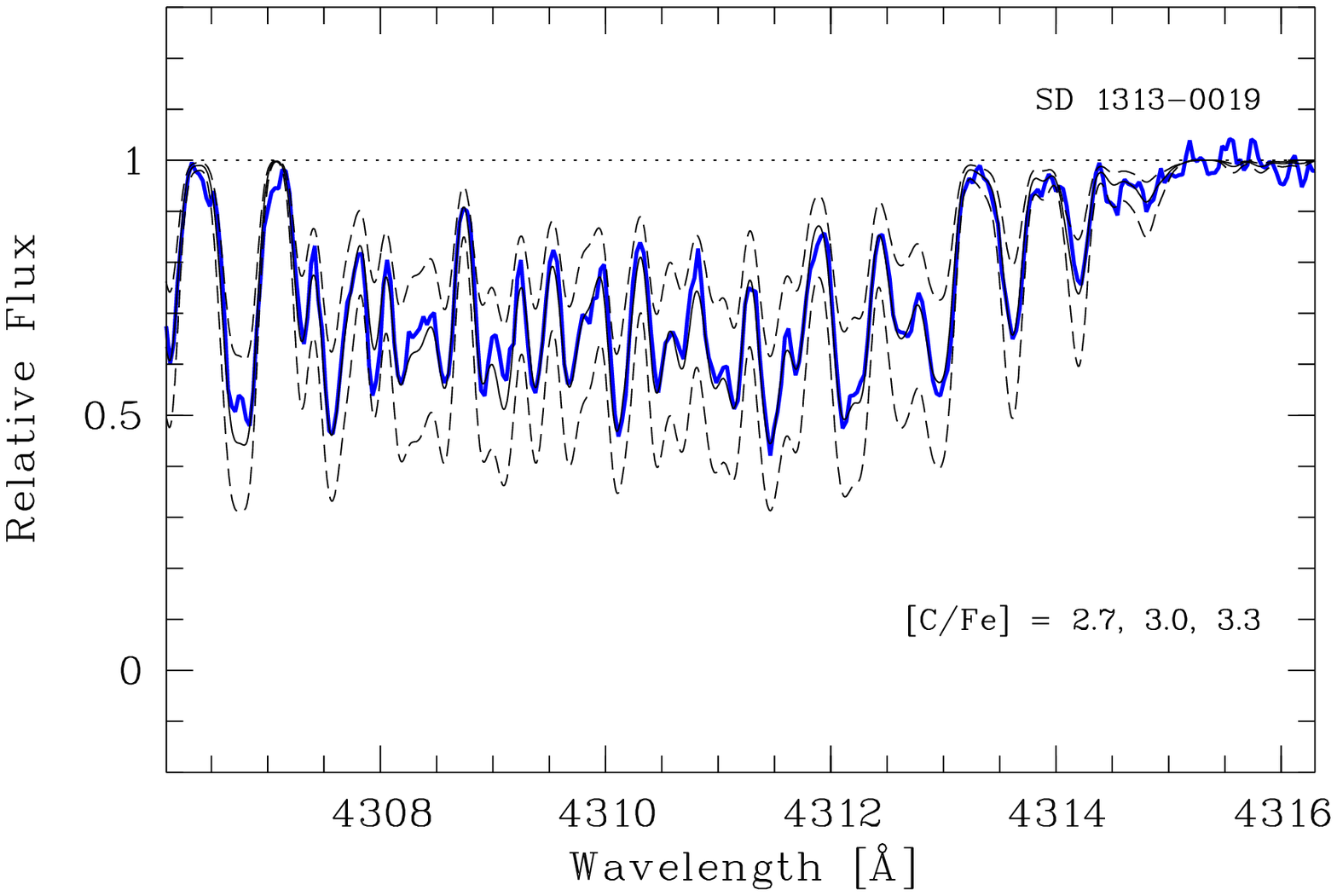}\\ \includegraphics[clip=true,width=7cm,
    bbllx=40, bblly=124, bburx=558,
    bbury=458]{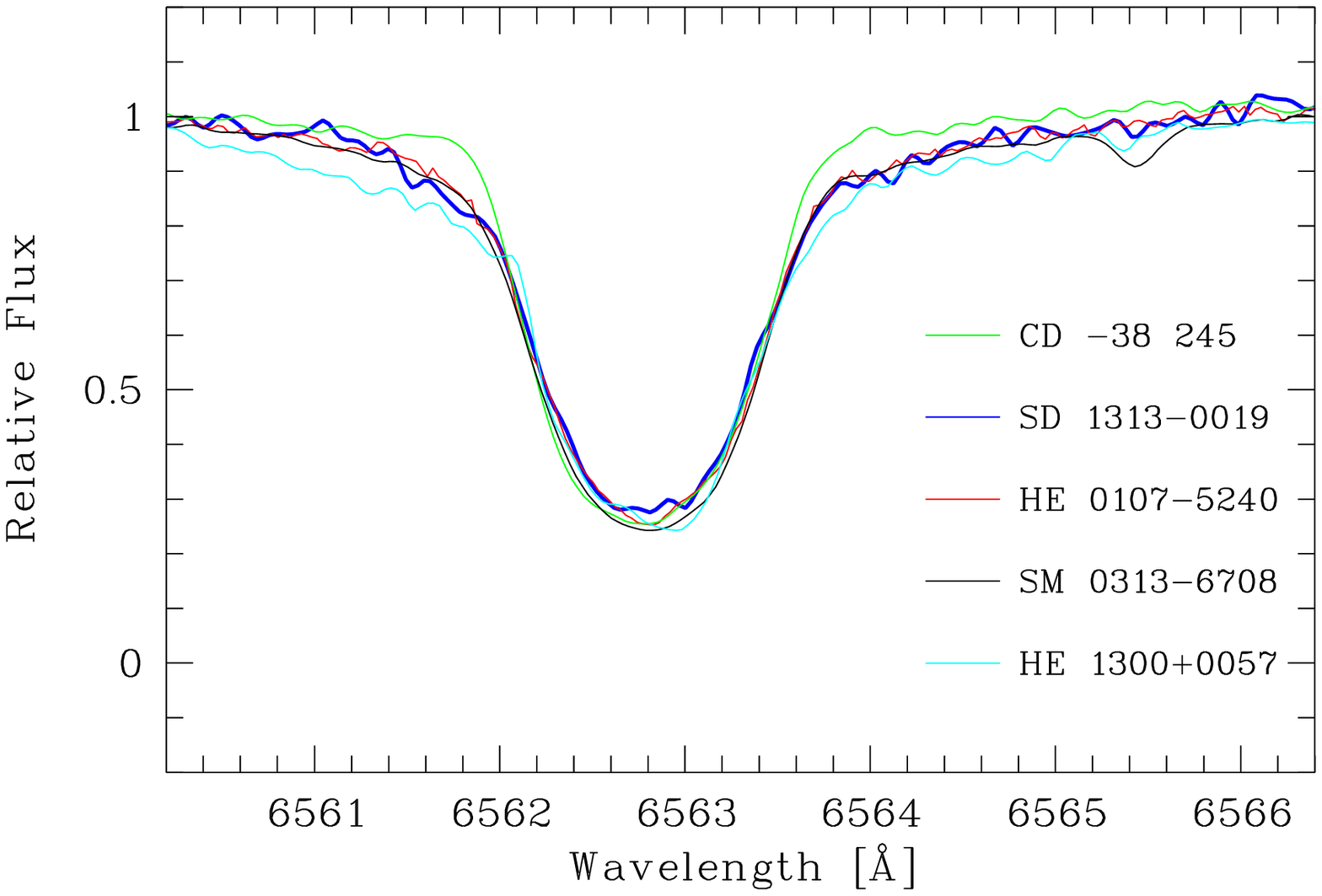}
 \figcaption{\label{specs} Portions of the Magellan/MIKE spectrum of
    SD~1313$-$0019 in comparison with other iron-poor stars near the
    Ca\,II\,K line at 3933\,{\AA} (top), around the Mg\,b lines around
    5180\,{\AA} (second panel), the G-band near 4313\,{\AA} (third
    panel) and H$\alpha$ line (bottom). Some absorption lines are
    indicated.} \end{center}
\end{figure}

Data reductions were carried out with the MIKE Carnegie Python
pipeline \citep{kelson03}\footnote{Available at
  http://obs.carnegiescience.edu/Code/python}.  The resulting $S/N$
per pixel is 45 at $\sim4700$\,{\AA}, 40 at $\sim5200$\,{\AA}, and 55
at $\sim6000$\,{\AA}. Radial velocity measurements yield
274.6\,km\,s$^{-1}$. This is consistent with the value of
$268\pm4$\,km\,s$^{-1}$ from the initial SEGUE spectrum
\citep{allendeprieto15} (spec-2901) from July 5, 2008 but different
from the value of $242\pm4$\,km\,s$^{-1}$ from the BOSS spectrum
(spec-7456) from March 11, 2014.  We checked our radial velocity zero
point with observations of G64$-$12 obtained in the same night as
SD~1313$-$0019. Its heliocentric radial velocity of
443.0\,km\,s$^{-1}$ agrees extremely well with literature values. As
already suggested by \citet{allendeprieto15}, our data supports
SD~1313$-$0019 being a member of a binary system, although additional
measurements are required to verify this.

In Figure~\ref{specs}, we show several representative portions of the
spectrum around the Ca\,II\,K line at 3933\,{\AA}, the G-bandhead at
4313\,{\AA}, and the Mg\,b lines at 5170\,{\AA}. We also show the
H$\alpha$ line in comparison with several other stars.

\section{Chemical abundance analysis}\label{sec:analysis}

\subsection{Stellar parameters}\label{stell_par_description}

We determined the stellar parameters spectroscopically, following the
procedure outlined in \citet{frebel13}. For that, we measured
equivalent widths, and carried out spectrum synthesis for blended
features and to determine upper limits of elements with no detected
lines using custom-made software \citep{casey14}. The equivalent
widths are presented in Table~\ref{Tab:Eqw}.  We used a 1D
plane-parallel model atmosphere with $\alpha$-enhancement 
\citep{castelli_kurucz} and a version of the MOOG analysis code that
accounts for Rayleigh scattering \citep{moog, sobeck11}. The
abundances are computed under the assumption of local thermodynamic
equilibrium (LTE). We derive an effective temperature of T$_{\rm
  eff}=5170$\,K, after applying the temperature correction (see Frebel
et al. 2013). We estimate an uncertainty of $\sim150$\,K given that we
only have 37 Fe\,I lines available.

This is somewhat cooler than the results of \citet{allendeprieto15}
who find values ranging from 5250 to 5670\,K, but adopt a value of
$\sim5300$\,K. To investigate reasons for these differences, we
visually compared the Balmer line strengths of SD~1313$-$0019 in our
spectrum to those of stars with similar temperature. As can be seen
in Figure~\ref{specs} (bottom panel), the shape of the H$\alpha$ line
of SD~1313$-$0019 agrees very well with those of HE~0107$-$5240
(T$_{\rm eff}=5100$\,K; \citealt{HE0107_ApJ}) and SM~0313$-$6708
(T$_{\rm eff}=5125$\,K; \citealt{keller14}) but not with that of the
warmer HE~1300+0057 (T$_{\rm eff}=5450$\,K; \citealt{frebel_he1300})
or the cooler CD$-$38~245 (T$_{\rm eff}=4800$\,K;
\citealt{HE0107_ApJ}).

In addition, we repeated our spectroscopic analysis to see if T$_{\rm
  eff}=5300$\,K can be supported. Within our equivalent width
measurement uncertainties, T$_{\rm eff}=5300$\,K also satisfies the
excitation equilibrium. We thus decided to adopt a temperature of
5200\,K, rather than 5100\,K following the Balmer line
comparison. Both are consistent with our initial spectroscopic result.

Since no Fe\,II lines were detected, we adopted a surface gravity from
an isochrone. We use one with $\mbox{[Fe/H]}=-3.0$ \citep{Y2_iso} but
one with $\mbox{[Fe/H]}=-2.0$, following the overall metallicity of
the star, would essentially give the same result. Our final stellar
parameters are T$_{\rm eff}=5200\pm150$\,K, $\log g=2.6\pm0.5$,
$v_{micr}=1.8\pm0.3$\,km\,s$^{-1}$ and $\mbox{[Fe/H]}=-5.0\pm0.1$.

\begin{deluxetable}{lrrrrr}
\tablecolumns{8}
\tablewidth{0pt}
\tabletypesize{\footnotesize}
\tabletypesize{\tiny}
\tablecaption{\label{Tab:Eqw} Equivalent width measurements}
\tablehead{
\colhead{Element}
& \colhead{$\lambda$}
&\colhead{$\chi$}
&\colhead{$\log gf$}&
\colhead{EW}&\colhead{$\log\epsilon$}\\
\colhead{}&\colhead{[{\AA}]}&\colhead{[eV]}&\colhead{[dex]}&\colhead{[m{\AA}]}&\colhead{[dex]}\\}
\startdata
Li    & 6707.7   & 0.00    & 0.170     & syn    & $<$0.80 \\
CH    & 4313     & \nodata & \nodata   & syn    & 6.44 \\
CH    & 4323     & \nodata & \nodata   & syn    & 6.34 \\
N (CN)\tablenotemark{a} & 3883 & \nodata & \nodata & syn & 6.29 \\
Na\,I  & 5895.92  & 0.00    & $-$0.19   & 9.7    & 1.61 \\
Mg\,I  & 3829.36  & 2.71    & $-$0.21   & 47.6   & 3.17 \\
Mg\,I  & 3832.30  & 2.71    & 0.27      & 66.9   & 3.00 \\
Mg\,I  & 3838.29  & 2.72    & 0.49      & 73.6   & 2.91 \\
Mg\,I  & 5183.60  & 2.72    & $-$0.24   & 42.4   & 3.03 \\
Mg\,I  & 5172.68  & 2.71    & $-$0.45   & 34.5   & 3.09 \\
Al\,I  & 3961.52  & 0.01    & $-$0.34   & 20.8   & 1.34 \\
Si\,I  & 3905.52  & \nodata & \nodata   & syn    & $<$2.72 \\
Ca\,I  & 4226.73  & 0.00    & 0.24      & 54.1   & 1.59 \\
Ca\,II & 3933.66  & 0.00    & 0.11      & 390.1  & 1.73 \\
Sc\,II & 4246.82  & \nodata & \nodata   & syn    & $-$1.54 \\
Ti\,II & 3759.29  & 0.61    & 0.28      & 50.5   & 0.33 \\
Ti\,II & 3761.32  & 0.57    & 0.18      & 47.5   & 0.32 \\
Mn\,I  & 4030.75  & 0.00    & $-$0.48   &$<$16.6 & $<$0.66 \\
Fe\,I  & 3878.57  & 0.09    & $-$1.38   & 45.2   & 2.54 \\
Fe\,I  & 3886.28  & 0.05    & $-$1.08   & 58.3   & 2.45 \\
Fe\,I  & 3887.05  & 0.91    & $-$1.14   & 13.8   & 2.47 \\
Fe\,I  & 3895.66  & 0.11    & $-$1.67   & 28.2   & 2.51 \\
Fe\,I  & 3899.71  & 0.09    & $-$1.52   & 40.1   & 2.57 \\
Fe\,I  & 3902.95  & 1.56    & $-$0.44   & 18.1   & 2.63 \\
Fe\,I  & 3920.26  & 0.12    & $-$1.73   & 32.2   & 2.66 \\
Fe\,I  & 3922.91  & 0.05    & $-$1.63   & 30.4   & 2.45 \\
Fe\,I  & 4045.81  & 1.49    & 0.28      & 45.7   & 2.41 \\
Fe\,I  & 4063.59  & 1.56    & 0.06      & 31.1   & 2.42 \\
Fe\,I  & 4071.74  & 1.61    & $-$0.01   & 27.3   & 2.47 \\
Fe\,I  & 4132.06  & 1.61    & $-$0.68   & 8.3    & 2.51 \\
Fe\,I  & 4202.03  & 1.49    & $-$0.69   & 12.0   & 2.56 \\
Fe\,I  & 4271.76  & 1.49    & $-$0.17   & 28.5   & 2.50 \\
Fe\,I  & 4325.76  & 1.61    & 0.01      & 28.9   & 2.46 \\
Fe\,I  & 4383.55  & 1.48    & 0.20      & 40.5   & 2.35 \\
Fe\,I  & 4404.75  & 1.56    & $-$0.15   & 20.2   & 2.35 \\
Fe\,I  & 4415.12  & 1.61    & $-$0.62   & 9.8    & 2.50 \\
Fe\,I  & 3840.44  & 0.99    & $-$0.50   & 44.1   & 2.64 \\
Fe\,I  & 3841.05  & 1.61    & $-$0.04   & 25.7   & 2.49 \\
Fe\,I  & 3856.37  & 0.05    & $-$1.28   & 59.2   & 2.67 \\
Fe\,I  & 3859.91  & 0.00    & $-$0.71   & 81.4   & 2.58 \\
Fe\,I  & 3490.57  & 0.05    & $-$1.11   & 56.0   & 2.56 \\
Fe\,I  & 3758.23  & 0.96    & $-$0.01   & 57.9   & 2.40 \\
Fe\,I  & 3763.79  & 0.99    & $-$0.22   & 52.4   & 2.53 \\
Fe\,I  & 3767.19  & 1.01    & $-$0.39   & 44.8   & 2.57 \\
Fe\,I  & 3787.88  & 1.01    & $-$0.84   & 18.5   & 2.45 \\
Fe\,I  & 3815.84  & 1.48    & 0.24      & 44.9   & 2.46 \\
Fe\,I  & 3820.43  & 0.86    & 0.16      & 70.5   & 2.39 \\
Fe\,I  & 3824.44  & 0.00    & $-$1.36   & 55.2   & 2.62 \\
Fe\,I  & 3825.88  & 0.91    & $-$0.02   & 63.6   & 2.46 \\
Fe\,I  & 3827.82  & 1.56    & 0.094     & 35.7   & 2.51 \\
Fe\,I  & 3618.77  & 0.99    & $-$0.00   & 59.5   & 2.55 \\
Fe\,I  & 3647.84  & 0.92    & $-$0.14   & 53.6   & 2.42 \\
Fe\,I  & 3719.94  & 0.00    & $-$0.42   & 82.6   & 2.37 \\
Fe\,I  & 3743.36  & 0.99    & $-$0.79   & 30.2   & 2.67 \\
Co\,I  & 3873.12  & 0.43    & $-$0.66  &$<$27.4 & $<$1.70 \\
Ni\,I  & 3858.30  & 0.42    & $-$0.95   & 18.1   & 1.58 \\
Ni\,I  & 3524.54  & 0.03    & 0.01      & 69.8   & 1.47 \\
Ni\,I  & 3807.14  & 0.42    & $-$1.22   & 20.0   & 1.91 \\
Zn\,I  & 4810.53  & 4.08    & $-$0.14   &$<$2.7  & $<$1.45 \\
Sr\,II & 4215.52  & 0.00    & $-$0.18   &$<$8.0  & $<-$2.41 \\
Ba\,II & 4554.03  & 0.00    &    syn    &$<$6.8  & $<-$2.60 \\
\enddata
\tablenotetext{a}{Holding the C abundance fixed at $\log\epsilon$(C) = 6.39.}
\end{deluxetable}

\subsection{Chemical abundances and measurement uncertainties}
\begin{figure*}[!ht]
 \begin{center}
  \includegraphics[clip=true,width=15cm,bbllx=5, bblly=15, bburx=630,
   bbury=477]{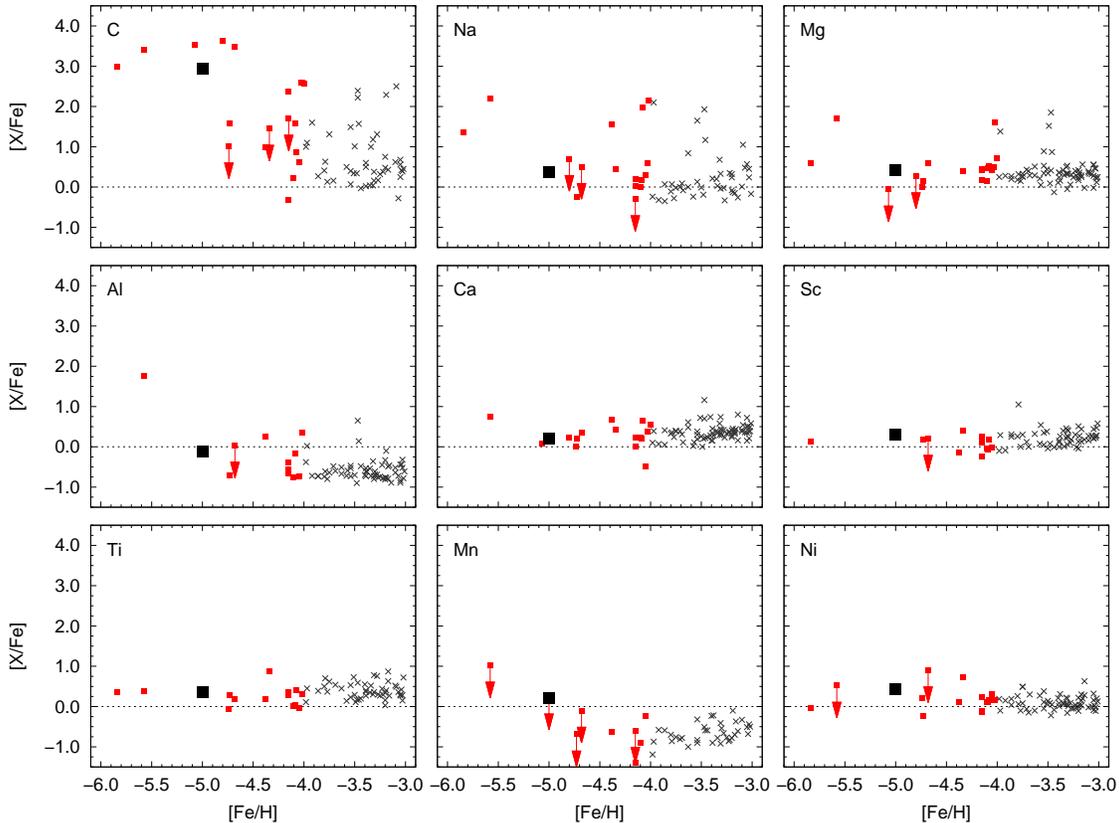} \figcaption{
     \label{abundplot} Abundance ratios ([X/Fe]) as a
     function of metallicity ([Fe/H]) for various elements detected in
     SD~1313$-$0019 (black square) and other halo stars (small red
     squares for stars with $\mbox{[Fe/H]}\le-4.0$, otherwise black
     crosses) (\citealt{yong13_II}, and references in \citealt{placco15}).}
 \end{center}
\end{figure*}

Chemical abundances were determined for 10 elements and upper limits
of an additional 7 elements. The final abundance ratios [X/Fe] are
calculated using the solar abundances of \citet{asplund09}. Our final
abundances are listed in Table~\ref{abund} and shown in
Figure~\ref{abundplot}.  We now briefly comment on each element
abundance.

Lithium was not detected. Its upper limit is A(Li) $< 0.8$ is consistent
with SD~1313$-$0019 being at the base of the red giant branch,
having begun to destroy its surface lithium abundances.

Carbon abundance was determined from the CH G-bandhead at 4313\,{\AA}
and an additional CH feature at 4323\,{\AA}, using the new linelist
by \citet{masseron14}. The best-fit synthetic spectrum is shown in
Figure~\ref{specs}. Due to the relatively unevolved nature of
SD~1313$-$0019, the carbon abundance does not need to be corrected for
evolutionary status \citep{placco14}.

We measured just the 5895\,{\AA} line of Na, with the other line at
5885\,{\AA} being distorted. There is excellent agreement
between the blue Mg triplet and green Mg~b lines. Only the Al
3961\,{\AA} line was used because the other available, at 3944\,{\AA}
is too heavily blended with CH. We obtained an
upper limit for Si from synthesis of the 3905\,{\AA} line, which is
blended with a molecular CH line.  Ca abundances were derived from
Ca\,I line at 4226\,{\AA} and the Ca\,II\,K line. The abundances
differ by 0.15\,dex, with a higher Ca\,II abundance. With
the Ca\,II\,K line having an equivalent width of 390\,m{\AA} we
regard it merely as confirmation of the Ca\,I
abundance. We also resolve at least one component of interstellar Ca
blueward of the Ca\,II\,K line. This was already noticed by
\citet{allendeprieto15}.

One weak Sc line at 4246\,{\AA} was tentatively detected and it is
blended with CH. Ti abundance was determined from two Ti lines,
yielding good agreement. Ni abundance was determined from three lines.
Upper limits for Mn, Co, Zn, Sr, and Ba were determined from the lines
at 4030\,{\AA}, 3873\,{\AA}, 4810\,{\AA}, 4077\,{\AA}, and
4554\,{\AA}, respectively.

Rgarding abundance uncertainties, we determined random
measurement uncertainties from the standard deviation when more than
one line was measured, although we adopt a nominal minimum value of
0.1\,dex. For elements with just one line, we assign uncertainties
based on abundance changes due to varying the continuum
placement. These uncertainties are listed in
Table~\ref{abund}. Concerning systematic uncertainties, we
re-determined the abundances after changing each of the stellar
parameters by their uncertainty while holding the others fixed. The
total uncertainties ($\sigma_{rand} + \sigma_{sys}$) for most elements
are about 0.2\,dex, and 0.4\,dex for [C/Fe] given its temperature
sensitivity.

\begin{deluxetable}{lrrrrr}
\tablewidth{0pc}
\tablecaption{\label{abund} Magellan/MIKE Chemical Abundances of SD~1313$-$0019}
\tablehead{
\colhead{Species} &
\colhead{$N$} &
\colhead{$\log\epsilon (\mbox{X})$} & \colhead{$\sigma$}&  \colhead{[X/H]}& \colhead{[X/Fe]}}
\startdata
Li (Syn.) & 1  & $<$0.80       & \nodata   & \nodata   & \nodata   \\
C (Syn.)  & 2  & 6.39          & 0.28      & $-$2.04   & 2.96      \\
N (Syn.)  & 1  & 6.29          & 0.40      & $-$1.54   & 3.46      \\
Na\,I      & 1  & 1.61          & 0.15      & $-$4.63   & 0.37      \\
Mg\,I      & 5  & 3.04          & 0.10      & $-$4.56   & 0.44      \\
Al\,I      & 1  & 1.34          & 0.30      & $-$5.11   & $-$0.11   \\
Si (Syn.) & 1  & $<$2.72       & \nodata   & $<-$4.79  & $<$0.21   \\
Ca\,I      & 1  & 1.59          & 0.10      & $-$4.75   & 0.25      \\
Ca\,II     & 1  & 1.73          & 0.20      & $-$4.61   & 0.39      \\
Sc (Syn.) & 1  & $-$1.54       & 0.30      & $-$4.69   & 0.31      \\
Ti\,II     & 2  & 0.33          & 0.10      & $-$4.62   & 0.37      \\
Mn\,I      & 1  & $<$0.66       & \nodata   & $<-$4.77  & $<$0.23   \\
Fe\,I      & 36 & 2.50          & 0.10      & $-$5.00   & 0.00      \\
Co\,I      & 1  & $<$1.70       & \nodata   & $<-$3.29  & $<$1.71   \\
Ni\,I      & 3  & 1.65          & 0.20      & $-$4.57   & 0.43      \\
Zn\,I      & 1  & $<$1.45       & \nodata   & $<-$3.11  & $<$1.89   \\
Sr\,II     & 1  & $<-$2.41      & \nodata   & $<-$5.28  & $<-$0.28  \\
Ba\,II     & 1  & $<-$2.60      & \nodata   & $<-$4.78  & $<$0.22   \\
\enddata
\end{deluxetable}

\section{Discussion}

\subsection{Origin of the abundance signature of SD~1313$-$0019 and other CEMP stars with $\mbox{[Fe/H]}\lesssim-5.0$}

SD~1313$-$0019 is a hyper iron-poor CEMP star with
$\mbox{[C/Fe]}=3.0$. Only four stars are currently known with lower
[Fe/H] values and only six other stars have [C/Fe] ratios of
$\mbox{[C/Fe]}\gtrsim3.0$. SD~1313$-$0019 thus contributes to the
100\% CEMP star fraction at the lowest iron abundances
\citep{placco14}. The carbon present in the birth gas clouds may have
originated in rotating massive stars which ejected strong stellar
winds (e.g., \citealt{meynet06}), in fallback supernovae (e.g.,
\citealt{tominaga14}), or by two supernovae (e.g.,
\citealt{limongi_he0107}). See \citet{fn15} for an extensive
discussion.

Alternatively, mass transfer from an erstwhile AGB binary companion
might be responsible for the observed high [C/Fe] ratio, given its
alleged radial velocity variation. Large amounts of s-process elements
would also be expected, such as $\mbox{[Ba/Fe]}>1.0$, typical for
s-process stars (e.g., \citealt{placco13}). This is, however,
currently ruled out by its upper limit of $\mbox{[Ba/Fe]}<0.2$.

\subsection{Constraining Population\,III star properties}

We now present an example analysis of the abundance pattern of
SD~1313$-$0019 using theoretical model predictions of supernova yields
from single non-rotating massive Pop\,III stars. While other yields
could also be used, due to space constraints, we restrict ourselves to
the \citet{heger10} fallback (S4) models\footnote{Accessible at
  http://starfit.org} which comprises 120 models with masses from 10
to 100\,M$_\odot$, covering explosion energies from $0.3 \times
10^{51}$\,erg to $10 \times 10^{51}$\,erg. For each energy, there are
models with different (fixed) mixing amounts (see Tables~8 and 9 of
\citealt{heger10}). The mass cut is fixed at all times and assumed to
take place at the base of the oxygen burning shell, where the
piston is placed.

We use their publicly available $\chi^2$ matching algorithm to
determine which model fits our abundances best. The fitting algorithm
assumes Sc and Cu are lower limits (generally, Sc is underproduced by
the yields), and it ignores Li, Cr, and Zn by default. For both Sc and
Cr, \citet{heger10} assume unaccounted for, additional production
sites. Any apparent discrepancy should thus be disregarded.

\begin{figure*}[!Ht]
 \begin{center}
  \includegraphics[clip=true,width=16cm, bbllx=5, bblly=23, bburx=505,
    bbury=338]{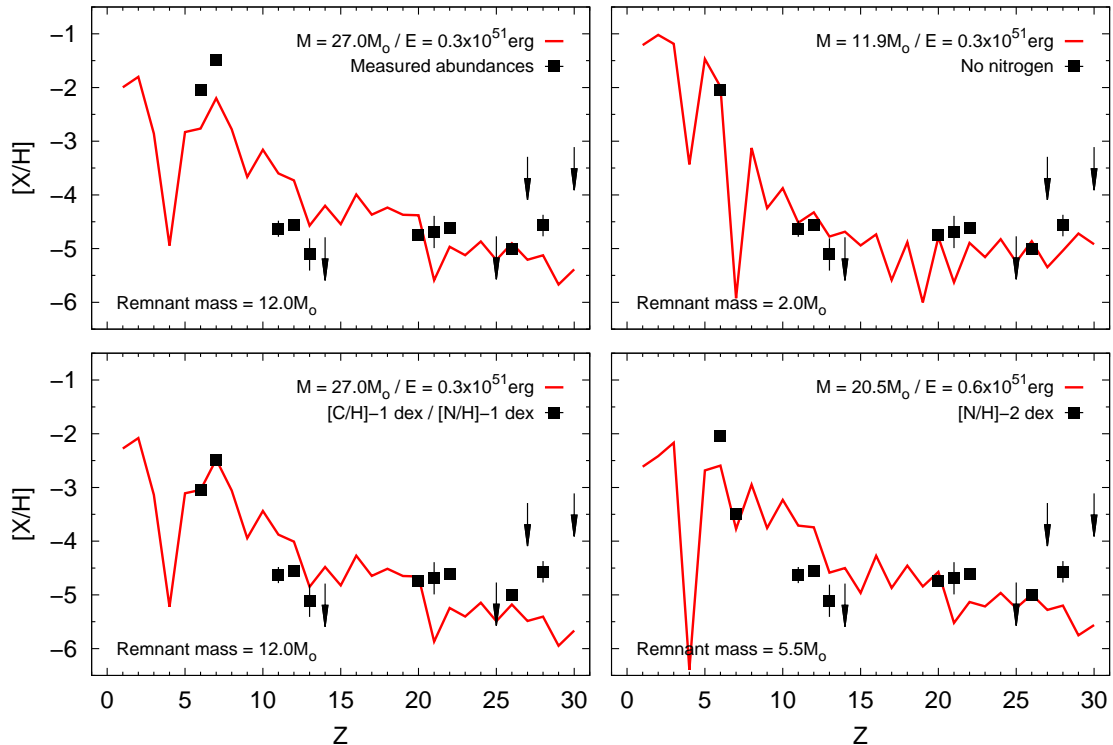}\\ \figcaption{\label{sn_fit} Abundance
    ratios [X/H] of SD~1313$-$0019 as a function of atomic number.
    Four variations on the abundances are shown (see upper right
    corners). The solid lines are the best fits for each abundance set
    using the S4 yields \citep{heger10}. The resulting model mass and
    energy are also shown in the upper right, and the remnant mass in
    the lower left. Arrows represent upper limits. See text for discussion.}
\end{center}
\end{figure*}

Results are shown in Figure~\ref{sn_fit}. Fitting the
abundance pattern (top left panel) yields a progenitor mass of
27\,M$_\odot$ and low explosion energy of 0.3 $\times 10^{51}$\,erg,
although the quality of the fit is only moderate. This could partly be
due to our high C and N abundances. Given that 3D effects for
abundances determined from molecular features (CH, NH, CN) can be
$>$1\,dex, leaves 1D LTE abundances overestimated compared to those
from 3D or time-averaged ($<$3D$>$) model atmospheres (e.g.,
\citealt{magic13}).

For example, for the CN feature, \citet{spite13} found 3D N abundances
to be as much as 2.4\,dex lower than 1D values. Therefore, we modified
our C and N abundances to investigate potential 3D effects. This
follows \citet{placco15} who demonstrate that the quality of the match
between the star's abundance pattern to the yields is
highly sensitive to the N abundance. To illustrate this issue, the
bottom left panel of Figure~\ref{sn_fit} shows the abundance
pattern with [C/H] and [N/H] each reduced by 1\,dex. There is no
change in the progenitor model but C and N are now fit well. There is
no change for the remaining elements.

To further illustrate the dependence of the best fit result on the N
abundance, we reduced [N/H] by 2\, dex but kept the original [C/H]
abundance. The progenitor properties changed only slightly;
20\,M$_{\odot}$ and 0.6 $\times 10^{51}$\,erg, and again, an improved
fit (bottom right panel). This test simulates the Spite et
al. suggestion of a reduced N abundance (when derived from CN). In
fact, a lower N value agrees with the upper limit derived from NH
lines at 3360\,{\AA}. For completeness, we also determine the best fit
after excluding the N abundance from the observed pattern (top right
panel). The progenitor mass is rather different then, with
12\,M$_{\odot}$, in line with findings by \citet{placco15}. 

We conclude that the progenitor of SD~1313$-$0019 had a mass of 20 to
30\,M$_{\odot}$, was of low explosion energy, and experienced a mixing
and fallback episode that led to a high [C/Fe] yield. Future work
exploring more details of the mixing and fallback supernovae or
alternate production scenarios (e.g., Limongi et al. 2003, Meynet et al.
2006) may be able to more closely reproduce the star's light element
abundances.

Another aspect to consider are the observed neutron-capture element
abundances and upper limits in the most iron-poor stars.  With
$\mbox{[Sr/H]} < -5.3$ and $\mbox{[Ba/H]} < -4.8$, SD~1313$-$0019
possesses extremely low neutron-capture abundances which are
among the lowest ever observed in halo stars (e.g., see Figure~8 in
\citealt{frebel14}). Interestingly, these abundance levels strongly
resemble those of the stars in the Segue\,1 ultra-faint dwarf galaxy
\citep{frebel14}. This might suggest that SD~1313$-$0019 formed in a
system not too dissimilar from Segue\,1 which is
characterized by its uniquely low neutron-capture element content.

More generally, the low Sr and Ba values of SD~1313$-$0019 extend the
decreasing abundance trends of halo stars with decreasing [Fe/H] down
to $\mbox{[Fe/H]}=-5.0$. The upper limits of [Sr/H] and [Ba/H] suggest
that neutron-capture elements were not made in every first generation
star. This could imply that neutron-capture processes are mass-
or rotation dependent and only happen in certain mass
ranges, such as the s-process in massive rotating low-metallicity
stars (\citealt{pignatari08}). Broadly, this would imply a
variety of progenitor masses because HE~1327$-$2326, with
$\mbox{[Fe/H]}=-5.6$ has a much larger Sr abundance
($\mbox{[Sr/H]}\sim-4.7$), in contrast to SD~11313$-$0019. However,
current abundance fitting of supernova yields only take into account
elements up to Zn, and those results suggest that the progenitors of
$\mbox{[Fe/H]}\lesssim-5.0$ stars have a relatively narrow mass range
\citep{placco15}. Additional iron-poor stars and especially more
extensive nucleosynthesis prediction, particularly for
neutron-capture element abundances, are needed to fully understand the
observed signatures. Nevertheless, it is becoming clear that
neutron-capture elements have the potential to strongly constrain
progenitor properties.

\subsection{The formation sites and mechanisms of the first low-mass stars}
The first Pop\,II stars formed in either $\sim 10^6$\,M$_\odot$
minihalos or $\sim 10^8$\,M$_\odot$ atomic cooling halos.
Either formation site is possible, but if the large observed [C/Fe]
ratios originated from low energy, faint supernovae, this suggests
that minihalos are the preferred formation site of CEMP stars
\citep{cooke14} or even externally enriched minihalos
\citep{smith15}. However, even if several $10^{51}$\,erg core-collapse
supernovae explode in a single minihalo, hydrodynamic simulations with
cosmological initial conditions have found that metals expelled by
Pop\,III stars are retained and second-generation stars form
\citep{ritter15}. Comparing the latter scenario to models of chemical
enrichment suggests that a majority of Pop\,III supernovae may have
provided the carbon-enhancement in the abundance signatures of the
most iron-poor stars (Ji et al. 2015, in prep).

Second-generation star formation may also have occured in chemically
homogeneous clusters \citep{blandhawthorn10, ritter15}, with two
low-mass stars from the same birth gas cloud ending in the stellar
halo. In this context, it is interesting to note the remarkably
similar abundance patterns of SD~1313$-$0019 and HE~0557$-$4840
\citep{he0557}. They have different C and N abundances, which could
potentially be brought into closer agreement by future NLTE and/or 3D
corrections. However, without dynamical information, it cannot be
excluded that these two stars formed from distinct supernovae had
produced similar abundance patterns.

Low-mass star formation may be facilitated by atomic carbon and
oxygen providing a cooling channel for the primordial gas to
sufficiently fragment \citep{brommnature}.  An alternate channel is
dust thermal cooling (e.g., \citealt{schneider12b}). Both
mechanisms can be tested with observational criteria
\citep{dtrans,ji14}.  The carbon abundance of SD~1313$-$0019 is
$\mbox{[C/H]} \sim -2.0$, easily satisfying the D$_{\rm trans}$
criterion for low-mass star formation through atomic line cooling
\citep{dtrans}. This star also has a low silicon abundance
$\mbox{[Si/H]} \lesssim -4.8$. If dust in the early universe was
predominantly silicates, it could not have formed from standard dust
cooling \citep{ji14}. Only if carbon dust is able to form or grain
growth is important, then dust cooling could have catalyzed
SD~1313$-$0019's formation \citep{chiaki15}.

\subsection{About future searches for CEMP stars with $\mbox{[Fe/H]}\lesssim-4.0$}

While \citet{allendeprieto15} found $\mbox{[Fe/H]}=-4.3$ for
SD~1313$-$0019 based on a medium-resolution spectrum ($R\sim2,000$),
we find a significantly lower metallicity even when taking the
effective temperature differences into account. Using 5378\,K, we find
$\mbox{[Fe/H]}=-4.82$. The Ca\,II\,K line, on which the
medium-resolution value was based, is blended with a double-peaked
carbon feature at 3935.5\,{\AA}. After smoothing our spectrum to a
comparable resolution, we recovered the quoted equivalent width of
640\,m{\AA} of \citet{allendeprieto15}. Our corresponding abundance is
$\mbox{[Ca/H]} \sim-4.0$ which agrees with their $\mbox{[Fe/H]}=-4.3$,
assuming $\mbox{[Ca/Fe]}\sim0.3$.

Since the most iron-poor stars are likely to be carbon-enhanced, they
display strong CH features throughout the blue part of the spectrum,
including near the Ca\,II\,K line. Examples include HE~0107$-$5240
\citep{HE0107_Nature} and SM0313$-$6708 \citep{keller14}, where the
equivalent width of the combined Ca\,II\,K-CH$_{3935.5}$ feature is
about twice that of the Ca\,II\,K line alone (see
Figure~\ref{specs}). In fact, the discrepancy between medium and
high-resolution spectra of nine of the most iron-poor stars varies
from $-0.5$ to $-1.75$\,dex.  It is thus impossible for this
population of stars to obtain an accurate [Fe/H] estimate from the
Ca\,K\,II unless this effect is taken into account.

With an independent C abundance (e.g., from the G-band, measurable in
medium-resolution spectra) and approximate stellar parameters, it
should be possible to predict the contribution of the CH feature at
3935.5\,{\AA} and subtract it from the equivalent width of the
Ca\,II\,K line. This will become particularly important for
both spectroscopic and narrow-band photometric surveys that contain
large amounts of fainter metal-poor stars that cannot be followed up
with high-resolution spectroscopy. Overestimating the iron abundances
of the most iron-poor stars due to this carbon contamination would
skew the shape of the metallicity distribution function at its tail
end, thus hampering our understanding of the formation of the earliest
long-lived stars as well as the stellar halo of Milky Way.

\acknowledgements{A.F. and A.P.J. are supported by NSF CAREER grant
  AST-1255160.  V.M.P. acknowledges partial support for this work from
  PHY 08-22648; Physics Frontier Center/{}Joint Institute for Nuclear
  Astrophysics (JINA) and PHY 14-30152; Physics Frontier Center/JINA
  Center for the Evolution of the Elements (JINA-CEE), awarded by the
  US National Science Foundation. This work made use of the NASA's
  Astrophysics Data System Bibliographic Services.}

\textit{Facilities:} \facility{Magellan-Clay (MIKE)}


\end{document}